\documentclass[%
reprint,
amsmath,amssymb,
aps,
]{revtex4-2}
\usepackage{graphicx}
\usepackage{rotating}
\usepackage{amsmath}
\usepackage{latexsym}
\usepackage{color}

\begin{document}
\title{Ranking dynamics of urban mobility}

\author{Hao Wang} \email{wanghaoo@stu.pku.edu.cn}
\affiliation{School of Urban Planning and Design, Peking University Shenzhen Graduate School, Shenzhen 518055, China}

\begin{abstract}
Human mobility, a pivotal aspect of urban dynamics, displays a profound and multifaceted relationship with urban sustainability. Despite considerable efforts analyzing mobility patterns over decades, the ranking dynamics of urban mobility has received limited attention. This study aims to contribute to the field by investigating changes in rank and size of hourly inflows to various locations across 60 Chinese cities throughout the day. We find that the rank-size distribution of hourly inflows over the course of the day is stable across cities. To uncover the microdynamics beneath the stable aggregate distribution amidst shifting location inflows, we analyzed consecutive-hour inflow size and ranking variations. Our findings reveal a dichotomy: locations with higher daily average inflow display a clear monotonic trend, with more pronounced increases or decreases in consecutive-hour inflow. In contrast, ranking variations exhibit a non-monotonic pattern, distinguished by the stability of not only the top and bottom rankings but also those in moderately-inflowed locations. Finally, we compare ranking dynamics across cities using a ranking metric, the rank turnover. The results advance our understanding of urban mobility dynamics, providing a basis for applications in urban planning and traffic engineering. 
\end{abstract}
\maketitle
\vspace{1cm}

\section{Introduction}
The rapid urbanization of our world is a defining trend of the 21st century, with projections indicating that nearly 70\% of the global population will reside in urban areas by 2050~\cite{Yabe2022,McDonnell2016,Brelsford2017}. This demographic shift underscores the critical need to comprehend the dynamics within urban systems. The quantitative analysis of human mobility stands at the forefront of urban studies~\cite{Simini2012,Yan2017,Gallotti2016,Alessandretti2020,Li2022,Lu2012}, crucial for a myriad of practical applications including epidemic containment~\cite{Jia2020,Chang2021,Vahedi2021}, traffic forecasting~\cite{Ortuzar2011}, and urban planning~\cite{Batty2008}. The advent of modern communication technologies has significantly enhanced the availability of large-scale, high-precision human mobility data (e.g., mobile phone records and social media data)~\cite{Barbosa2018}. This development offers unprecedented opportunities to study human mobility patterns, including the statistical properties of individual human mobility~\cite{Brockmann2006,Gonzalez2008,Song2010}, community detection based on population flow~\cite{Zhang2023}, and the overarching characteristics of urban mobility~\cite{Louail2015,Bassolas2019,Zhao2024,Xu2023}. Although the significant insights revealed by recent human mobility research, there has been relatively little research on the ranking dynamics of urban mobility. 

Given the world's addiction to ranking, which impacts areas from scientists, journals, and universities to countries~\cite{Blumm2012,Michel2011}, assessing the ranking dynamics associated with urban mobility is particularly important. Traditionally, researchers primarily use the heavy-tailed decay of size with rank, commonly known as Zipf’s law, to characterize the statistical properties of rankings~\cite{Newman2005}. This law has been observed in various contexts, such as the ranking of words and phrases by frequency of use~\cite{Zipf1949}, cities by population~\cite{Rosen1980}, companies by size~\cite{Axtell2001,Stanley1996}, and earthquakes by magnitude~\cite{Sornette1996}. However, these observations are typically analyzed at a single instant of time, ignoring how rankings change over time~\cite{Blumm2012,Batty2006,Iniguez2022,Verbavatz2020}. For example, the inflows to each location rise and fall many times throughout the day.

Recently, some studies have begun to focus on the dynamics of ranking processes. For instance, Blumm {\it et al.} examined the universal features of ranking dynamics and developed a continuum theory predicting the existence of three dynamically distinct phases~\cite{Blumm2012}. Furthermore, Iñiguez {\it et al.} explored the dynamics of 30 rankings across various systems, revealing a continuum from high-rank to low-rank stability, and demonstrated that observed rank stability patterns can be explained by fundamental mechanisms involving element displacement and replacement~\cite{Iniguez2022}. Moreover, Somin {\it et al.} investigated the differences between the dynamics of node degrees and node ranks, finding that preferential principles do not apply to ranking changes, which instead follow a non-monotonic, inverse U-shaped curve~\cite{Somin2022}. However, the research on the ranking dynamics of urban mobility is still lacking. Unraveling the ranking dynamics of urban mobility can deepen our understanding of complex urban system dynamics, empowering policymakers and regulators with targeted intervention strategies to adjust system properties.

We aim to bridge this crucial gap by examining the changes in the size and ranking of hourly inflows at each location throughout the day. This analysis leverages data from approximately 90 million mobile phone users across 60 Chinese cities in August and November 2019. We first investigate the rank-size distribution of hourly inflows over the course of the day. Subsequently, we analyze the correlation between hour-to-hour inflow variations and the daily average inflow for each location. Then, we explore how changes in inflow rankings across consecutive hours relate to the daily average inflow for each location. Finally, we compare the ranking dynamics of mobility across cities.

\section{Data}
The mobility data utilized in this work are derived from an aggregated and anonymized mobile phone dataset provided by a Chinese telecommunications operator~\cite{Zhao2024}. The operator divided the city into a grid system of 0.005° $\times$ 0.005° (approximately 500m $\times$ 500m at the equator). This dataset encapsulates the hourly number of trips between different grids during August and November 2019, encompassing approximately 90 million residents across 60 Chinese cities. Each record includes the date, the hour, the longitude and latitude coordinates of both the origin and destination grids, as well as the number of trips made between them. To unravel the intricate ranking dynamics of urban mobility throughout the course of the day, we calculated the average hourly trips between grids from the mobile phone dataset for 25 typical weekdays (Tuesday to Thursday). By focusing on this particular weekday window, we aimed to minimize the influence of outliers caused by weekends or holidays, which might skew our understanding of typical weekday mobility patterns. Fig. 1a illustrates the hourly variation in trip volumes, revealing a consistent pattern across cities. Specifically, two pronounced peak periods emerge: the morning peak, which culminates between 7am and 8am, and the evening peak, which surges between 5pm and 6pm.

Additionally, the mobile phone dataset was gathered based on the administrative regions of individual cities, disregarding socioeconomic and morphological factors within those regions. To achieve a consistent and harmonized delineation of cities, it is necessary to identify urban areas that transcend the administrative divisions set by governmental authorities. Here, we adopted the global human settlement dataset, which defines an urban area as a contiguous region of 1 km$^2$ with either a population density of at least 1,500 individuals per km$^2$ or a built-up land coverage of at least 50\% per km$^2$, and the entire designated urban area must have a minimum total population of 50,000 inhabitants~\cite{Melchiorri2024}. Figs. 1b-c provide a detailed depiction of the identified urban area. It can be observed that a majority of the grids with significant inflows are located within the urban area, indicating a strong consistency between the mobile phone data and the urban area. This consistency, in turn, serves as an indirect confirmation of the reasonableness of the urban area. Through the aforementioned data processing, we obtained the hourly inflow volume for each grid within the urban area during a typical weekday (Figs. 1d-e).

\section{Method}
\subsection{Zipf’s law}
Here, we set out to verify whether the inflow volume adheres to Zipf’s law, a renowned methodology for examining the relationship between ranks and sizes in distributions. According to Zipf’s law~\cite{Zipf1949}, the size of entities—in this context, the inflow volume of grids—follows an inverse relationship with their rank, meaning that large inflows are rare, whereas smaller inflows are more abundant. Formally, Zipf’s law for the rank-size distribution can be expressed as:
\begin{equation}
	P(r) \sim r^{-\upsilon},
\end{equation}
where $P(r)$ represents the inflow volume of the grid occupying the $r$-th rank, with grids sorted by their inflow in decreasing order. The symbol $\upsilon$ denotes Zipf’s exponent. When $\upsilon$ equals 1, it represents the ideal Zipf’s law, which stipulates that the second largest grid possesses half the inflow volume of the largest, and this proportionality continues for subsequent ranks. As the value of $\upsilon$ increases, the distribution of grid sizes in terms of inflow volume becomes increasingly uneven, indicating that a few large grids dominate, capturing a substantial portion of the total inflows, while the majority of grids remain relatively small in comparison. Conversely, when $\upsilon$ approaches 0, it signifies that the inflow volumes of all grids tend to converge, leading to a highly uniform distribution~\cite{Arshad2018}.

\subsection{Rank turnover}
Ranking lists generally have a fixed size $N_0$, such as the top 100, enabling elements to enter or exit the list at any of the $T$ observations, $t = 0$, $\cdots$, $T - 1$, thereby facilitating the measurement of the flux of elements across rank boundaries~\cite{Gerlach2016}. The rank turnover~\cite{Iniguez2022} at time $t$ is defined as 
\begin{equation}
	\Phi_t = \frac{N_t}{N_0},
\end{equation}
where $N_t$ represents the cumulative number of unique elements that have appeared in the ranking list until time $t$, and $N_0$ is the fixed size of the ranking list. Rank turnover is a monotonic increasing function that quantifies the pace at which new elements enter the ranking list over time. By averaging the rank turnover over the entire observation period, we obtain the average turnover rate
\begin{equation}
	\bar{\Phi} = \frac{\Phi_{T-1} - \Phi_{0}}{T - 1},
\end{equation}
Here, $\Phi_{T-1}$ represents the rank turnover at the final observation time $T-1$, while $\Phi_{0}$ is typically 1. The average turnover rate thus captures the overall rate of change in the list’s composition over time. A higher value of $\bar{\Phi}$ signifies a faster pace of change in the ranking list. Conversely, a lower average turnover rate suggests a more stable ranking, where the same set of elements tends to dominate over an extended period.

\section{Results}

\subsection{Rank-size distribution for mobility inflows}
We commence our analysis by exploring the rank-size distribution of hourly inflows across various locations throughout the day. Fig. 2 presents the results for the top 100 hotspots in the sampled cities, revealing that the inflows, when sorted in descending order, conform to Zipf’s law with their ranks, and notably, the Zipf’s exponent remains stable across different hours within a day. This outcome underscores the hierarchical organization of urban mobility flows, where a certain degree of concentration is expected. However, the consistent observation that the Zipf exponents across all cities are less than 1 signifies a deviation from the ideal Zipf’s law ($\upsilon=1$). This deviation indicates a relatively less pronounced concentration of inflows at the top-ranked locations.

Beyond grasping the stable Zipf distribution of urban mobility inflows lies the problem of temporal evolution as location rankings continually shift. Studies have revealed that the evident macro-stability of Zipf distribution over time can mask a volatile and turbulent microdynamics, where locations swiftly alter their inflow rank-order even as their aggregate distribution appears quite stable~\cite{Batty2006,Barthelemy2019}. The constant shuffling of rankings, with many locations entering and exiting the top 100 throughout the day, is completely hidden by Fig. 2. To illustrate, in Beijing, a remarkable 347 distinct locations make it into the top 100 hotspots over a 24-hour span. However, only 8 of these locations consistently remain in the top 100 throughout the day. Therefore, it becomes clear that the observed stability of the Zipf distribution, while useful for certain analytical purposes, fails to capture the full extent of the microdynamics associated with urban mobility.

\subsection{Time variation of mobility inflows}

To capture these microdynamics, we analyze the relationship between consecutive-hour inflow variation $\Delta P$ and daily average inflow $\bar{P}$ for each location, as depicted in Fig. 3. One the one hand, we observe a clear monotonic trend, where locations with higher daily average inflow $\bar{P}$ exhibit more significant increases or decreases in consecutive-hour inflow $\Delta P$, echoing the preferential attachment~\cite{Jeong2003} and detachment~\cite{Brot2013} patterns in complex networks. On the other hand, we observe that $\Delta P$ fluctuates symmetrically around $\Delta P=0$. Hence, the inflow of each location has a comparable probability of moving upwards or downwards in the next time step, which is similar to the changes in systems such as word usage, Medicare, and market capitalization~\cite{Blumm2012}. However, we know that cities have their own rhythms~\cite{Louail2014}. When we break down the hourly variations, we observe a remarkable asymmetry for high $\bar{P}$ values (Fig. 4), indicating that high-inflow locations tend to exhibit an enhanced tendency to increase their inflow during certain hours, or conversely, to drop in inflow during other hours. In summary, this result implies that the symmetrical fluctuation of $\Delta P$ throughout the day in terms of location inflow stems from the superposition of the asymmetry in consecutive hourly variations and the unique rhythm inherent to urban mobility.

Furthermore, we investigate the relationship between the diurnal dispersion $\sigma_{\Delta P}$ of inflow variations $\Delta P$ and the daily average inflow $\bar{P}$ across locations, as depicted in Fig. 5. We find that $\sigma_{\Delta P}$ as a function of $\bar{P}$ adheres to a power-law relationship
\begin{equation}
	\sigma_{\Delta P} \sim \bar{P}^{\alpha},
\end{equation}
where $0.84 \leq \alpha \leq 0.95$. The fact that $\alpha \leq 1$ indicates that relative changes are more subdued for items ranking at the top, a phenomenon well-established in economic context, where larger companies exhibit lower volatility compared to their smaller counterparts~\cite{Mantegna1995}. This sublinear characteristic may contribute to the stability of high-ranking locations, as they are less prone to extreme fluctuations in inflow.

\subsection{Time variation of mobility inflow ranking}
Our next analysis explores the relationship between changes in inflow ranking $\Delta R$ across consecutive hours and the daily average inflow $\bar{P}$ for each location. Ranking is a collective measure: it depends not only on the inflow of the location itself, but also on the inflow of all other locations. Fig. 6 illustrates a surface plot of $\Delta R$ versus $\bar{P}$, exhibiting a non-monotonic pattern distinct from the variations in $\Delta P$ (as seen in Fig. 3), featuring stable top and bottom rankings. Specifically, locations with extremely high average flows atop the ranking remain stable due to their substantial flow base, requiring substantial shifts to alter positions significantly. Conversely, sites with minimal average flows at the bottom also exhibit stability, as minor flow changes hardly impact their rankings given their low flow baseline. This stability in the top and bottom rankings has also been observed across various other systems~\cite{Iniguez2022}.

However, interestingly, in the urban mobility system, certain cities exhibit a group of locations with moderate inflows that have relatively stable rankings (Figs. 6a-f). To uncover the reasons behind the stable rankings in moderately-inflowed locations, we measure the inflow distribution across all locations (Fig. 7). Our findings reveal that, for these cities, the density of locations falling within this moderate inflow range is notably sparse (Figs. 7a-f). This sparsity of location numbers implies that even if individual sites experience fluctuations in their inflow volumes, the overall impact on the ranking structure remains limited. This understanding provides valuable insights into how urban locations maintain their rankings amidst dynamic changes in urban mobility.

\subsection{Comparison of ranking dynamics across cities}
To compare the ranking dynamics across cities, inspired by the work of Iñiguez {\it et al.}\cite{Iniguez2022}, we narrow down the hourly observations for each city to focus on the top 100 locations. This approach enables locations to enter or exit the top 100 list at any given hour of the day, allowing us to quantify the movement of these locations across rank boundaries. By using the rank turnover metric $\Phi_t$ (see Methods), we measure the pace of changes in the ranking of the top 100 locations across these cities, as depicted in Fig. 8a. Notably, Beijing stands out as the most volatile city, exhibiting an average turnover rate $\bar{\Phi}$ of 0.107 (Fig. 8b). In stark contrast, Taizhou displays the least volatility, with an average turnover rate of just 0.01 (Fig. 8b). This finding underscores that in Beijing, the top 100 locations attracting substantial inflows undergo significant shifts throughout the day. In stark contrast, Taizhou’s top 100 popular destinations maintain their status consistently, regardless of the hour, demonstrating remarkable stability throughout the day. 

Next, we turn our attention to exploring the link between the average turnover rate and the average daily travel distance to gain further insights into urban dynamics. The average travel distance serves as a crucial indicator of mobility efficiency, fuel consumption, and CO$_2$ emissions. Our analysis reveals that the average turnover rate exhibits a strong positive correlation with the average travel distance (Fig. 8c). Specifically, the faster the pace of change in the ranking of inflows throughout the day, the greater the average travel distance tends to be. This is likely because people need to move between different functional areas at various times of the day, such as congregating in commercial and office districts during the day and returning to residential areas at night. The rapid fluctuations in inflow rankings suggest a pronounced spatial segregation among these functional areas, which in turn results in longer average travel distances, and vice versa.

\section{Conclusion and discussion}
The past decade has witnessed significant efforts to uncover and understand intricate human mobility patterns~\cite{Brockmann2006,Gonzalez2008,Song2010}. Previous studies have primarily focused on basic mobility aspects, such as jump length distributions and inter-location flows, resulting in models that replicate these patterns with remarkable accuracy. However, a crucial yet frequently overlooked aspect of human mobility pertains to the ranking dynamics of urban mobility. This paper endeavors to redress this critical gap by delving into the intraday variations in the size and rank of hourly inflows to diverse locations across 60 Chinese cities. We can capture the diurnal ranking dynamics within cities, providing a comprehensive view of urban mobility and insights into the rhythm of urban life.

We initially scrutinized the diurnal rank-size distribution of hourly inflows across locations, noting stable Zipf’s exponents for top 100 spots within a day. However, studies reveal that macro-stability conceals turbulent micro-dynamics~\cite{Batty2006,Barthelemy2019}, with locations rapidly shuffling ranks amidst an ostensibly stable aggregate inflow distribution. Driven by this, we delved into consecutive-hour variations in inflow size and ranking to unravel the intricate dynamics. The variations in consecutive-hour inflow sizes exhibit a clear monotonic pattern, wherein locations with higher daily average inflows demonstrate more pronounced hourly fluctuations. Furthermore, the variations are symmetric around zero, reflecting balanced surges and declines. We further found that this symmetry stems from the interplay of asymmetric hourly changes and the unique urban mobility rhythm. Regarding consecutive-hour inflow ranking variations, a non-monotonic pattern arises, with stable top and bottom rankings. Notably, certain cities exhibit resilient stability in moderately-inflowed locations due to the sparse density of locations that fall within this inflow range. This scarcity limits the overall impact of individual fluctuations on the ranking structure. Finally, we compared the ranking dynamics of urban mobility across different cities and found that the faster the pace of change in the ranking of inflows throughout the day, the greater the average travel distance tends to be.

Our findings extend the current literature on ranking dynamics. Blumm {\it et al.}, in their investigation of size fluctuations, identified two distinct systems~\cite{Blumm2012}. One system demonstrates symmetric fluctuations around zero in size changes between adjacent time periods, suggesting a comparable likelihood for items to increase or decrease in size. The other system exhibits a notable asymmetry about zero, indicating that high-scoring items have a heightened tendency to experience a decrease in score. Our study constructs a bridge between these two different systems, revealing that the symmetry observed in variations of consecutive-hour inflow sizes originates from the interaction of asymmetric hourly changes and the rhythm of urban mobility. In addition, previous studies have shown that although score distributions differ across systems, ranking lists for less open systems exhibit similar stability features, specifically stable top and bottom ranks~\cite{Iniguez2022}. Our study contributes to this existing literature by revealing that, in less open systems such as urban mobility, some cities not only exhibit stable top and bottom ranks but also show relatively stable rankings in areas of moderate inflow. This knowledge can empower decision-makers to anticipate and respond to changes in mobility patterns, enhancing the overall resilience and livability of urban environments.

In this study, we focused on examining the characteristics of rank dynamics associated with urban mobility. However, we acknowledge that our understanding of the mechanisms that give rise to these features remains limited. To gain a deeper understanding of these mechanisms and further construct models that can accurately reproduce these features, we need to dig deeper into the fundamental principles underlying ranking dynamics. Furthermore, our analysis of the correlation between the rate of ranking shifts and the average travel distance within urban regions was merely preliminary. Future research should delve deeper into the connections between ranking dynamics in urban mobility and other key urban indicators, such as hotspot flow matrix~\cite{Louail2015} and flow hierarchy~\cite{Bassolas2019}. Such explorations will enable us to grasp the intricacies and varied facets of urban mobility with greater clarity and depth.

\section*{Competing interests}
\noindent
The author declare no competing financial interests.

\begin{figure*}
	\centering
	{\includegraphics[width=\linewidth]{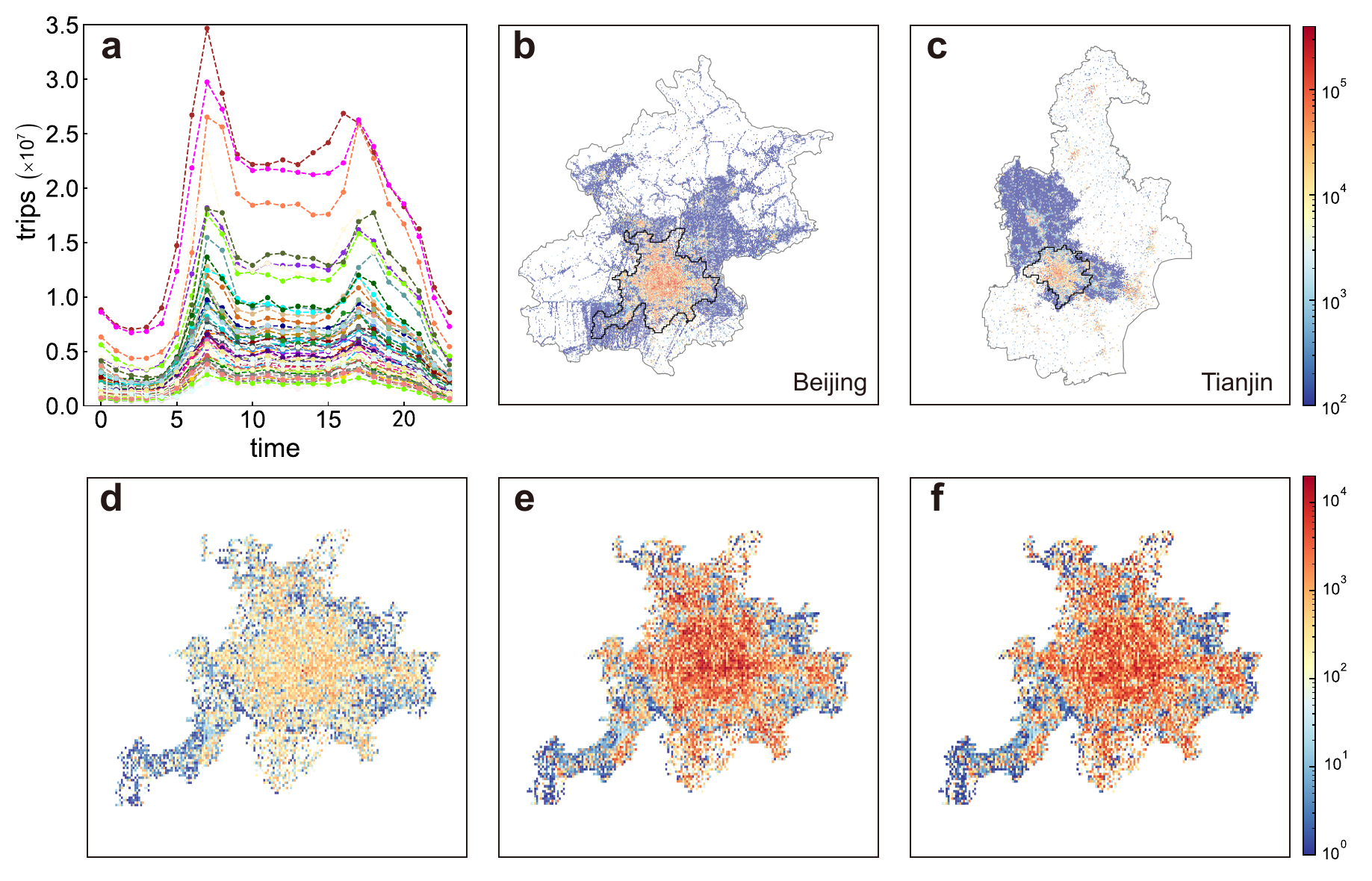}}
	\caption{{\bf Illustration of the mobility data.} 
		(a) Time evolution of the number of trips per hour during a typical weekday. Each city is distinguished by a unique color, where each point signifies the number of trips recorded within that hour.
		(b-c) Comparison of the urban area and the inflow spatial distribution in (b) Beijing and (c) Tianjin. The administrative boundary is denoted by the grey line, while the urban area is outlined in black. Each grid is color-coded to represent the intensity of the total daily inflow volume.
		(d-f) Spatial distribution maps of inflow volume for each location in Beijing at (d) 3am, (e) 7am, and (f) 5pm. The color of each grid is proportional to the inflow volume within that hour for that grid.
	}
	\label{fig1}
\end{figure*}

\begin{figure*}
	\centering
	{\includegraphics[width= \linewidth]{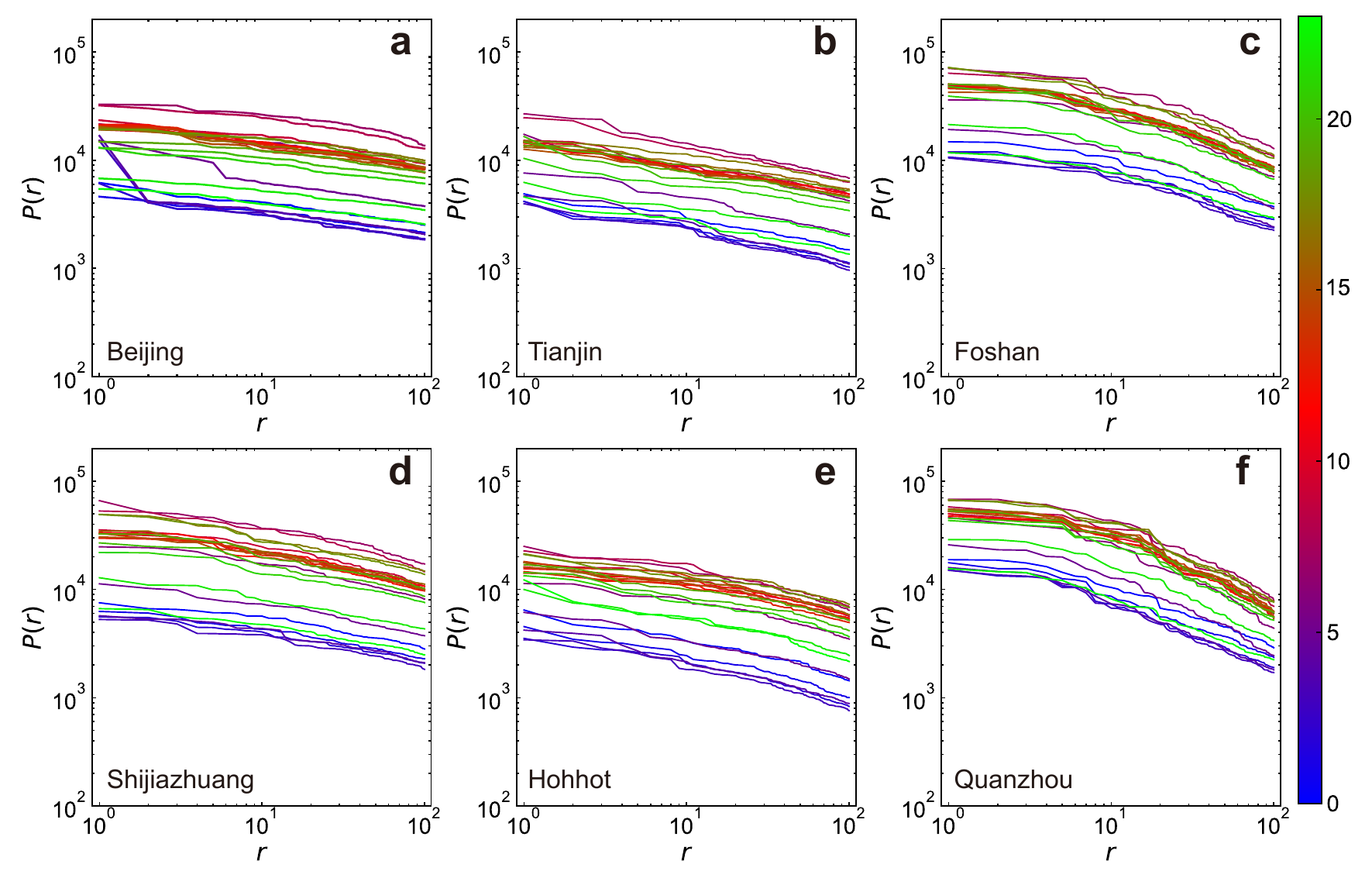}}
	\caption{{\bf Zipf plots of the top 100 locations in six sampled cities.}
		(a) Beijing, (b) Tianjin, (c) Foshan, (d) Shijiazhuang, (e) Hohhot, and (f) Quanzhou. Line colors denote different hours of the day, with green and blue signifying early morning and late night, and yellow and red indicating midday.
	}
	\label{fig2}
\end{figure*}

\begin{figure*}
	\centering
	{\includegraphics[width=\linewidth]{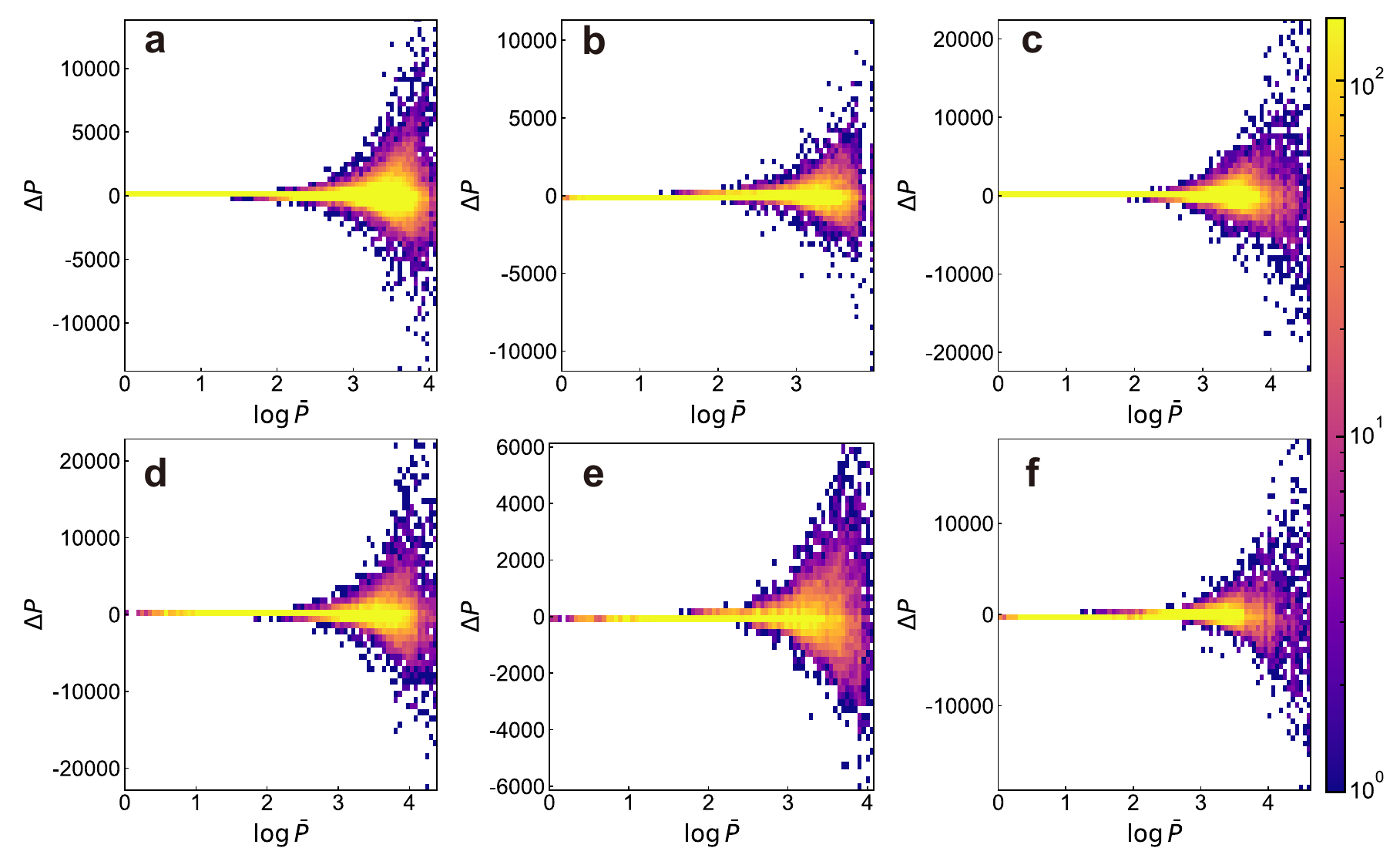}}
	\caption{{\bf The variation in consecutive-hour inflow as a function of daily average inflow.} 
		(a-f) Surface plots show $\Delta P$ trends with $\bar{P}$ in six sampled cities: (a) Beijing, (b) Tianjin, (c) Foshan, (d) Shijiazhuang, (e) Hohhot, and (f) Quanzhou. 
	}
	\label{fig3}
\end{figure*}

\begin{figure*}
	\centering
	\includegraphics[width=\linewidth]{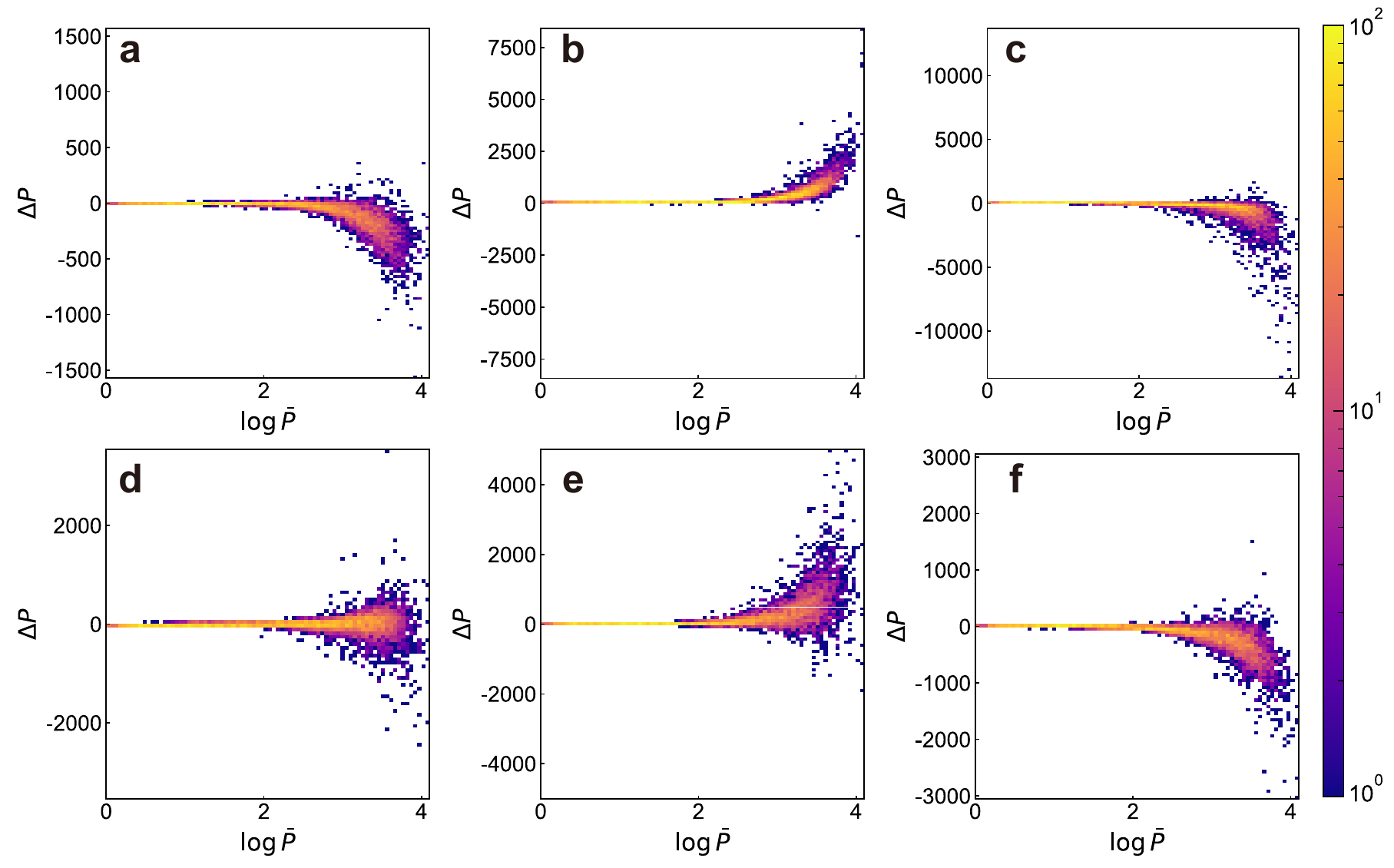}
	\caption{{\bf Surface plots illustrate the trends of $\Delta P$ with respect to $\bar{P}$ in Beijing at different hours.}
		(a) 00:00, (b) 04:00, (c) 08:00, (d) 12:00, (e) 16:00, and (f) 20:00.
	}
	\label{fig4}
\end{figure*}

\begin{figure*}
	\centering
	{\includegraphics[width=\linewidth]{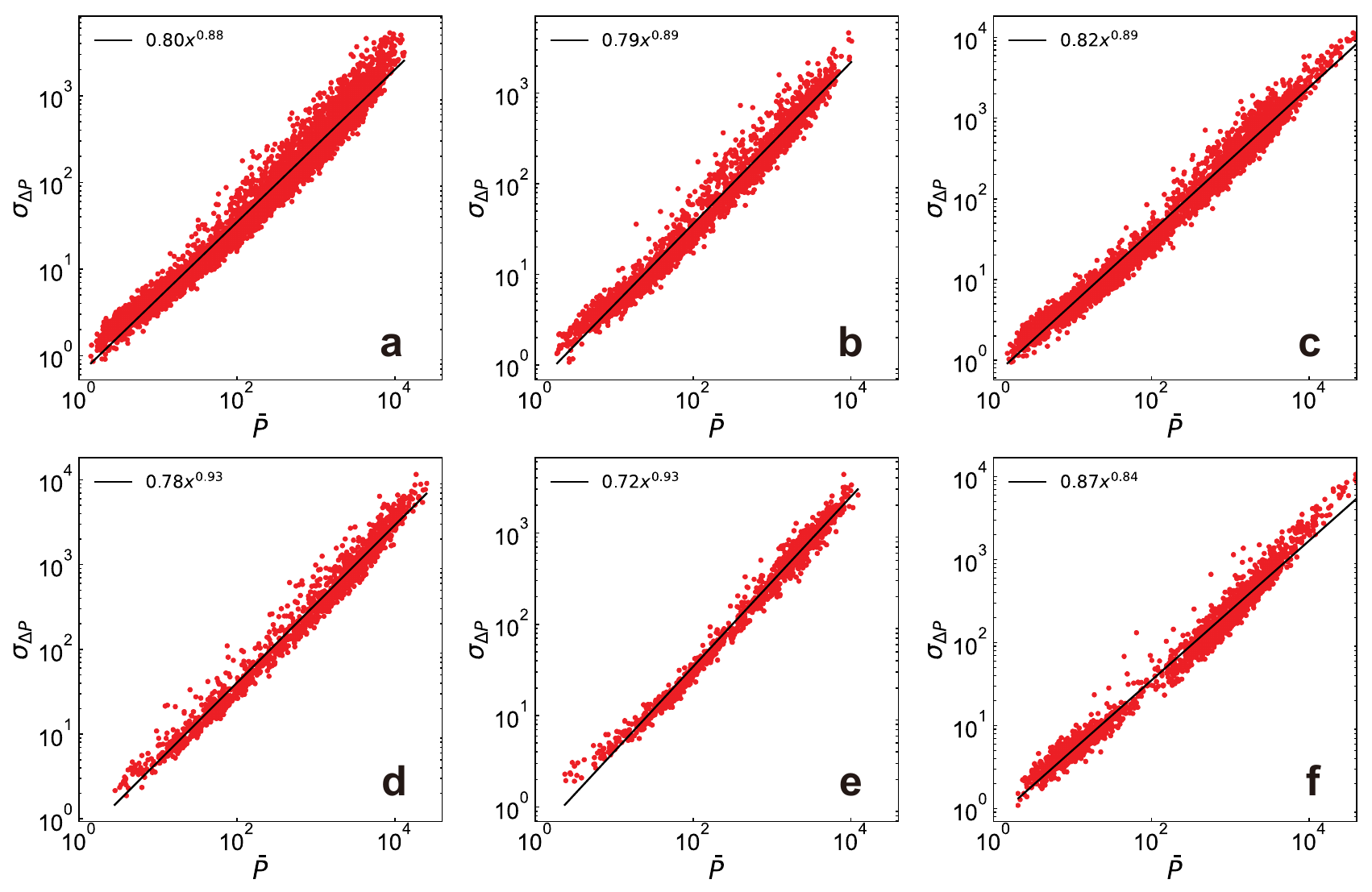}}
	\caption{{\bf $\sigma_{\Delta P}$ as a function of $\bar{P}$ in six sampled cities.}
		(a) Beijing, (b) Tianjin, (c) Foshan, (d) Shijiazhuang, (e) Hohhot, and (f) Quanzhou.
	}
	\label{fig5}
\end{figure*}

\begin{figure*}
	\centering
	{\includegraphics[width=\linewidth]{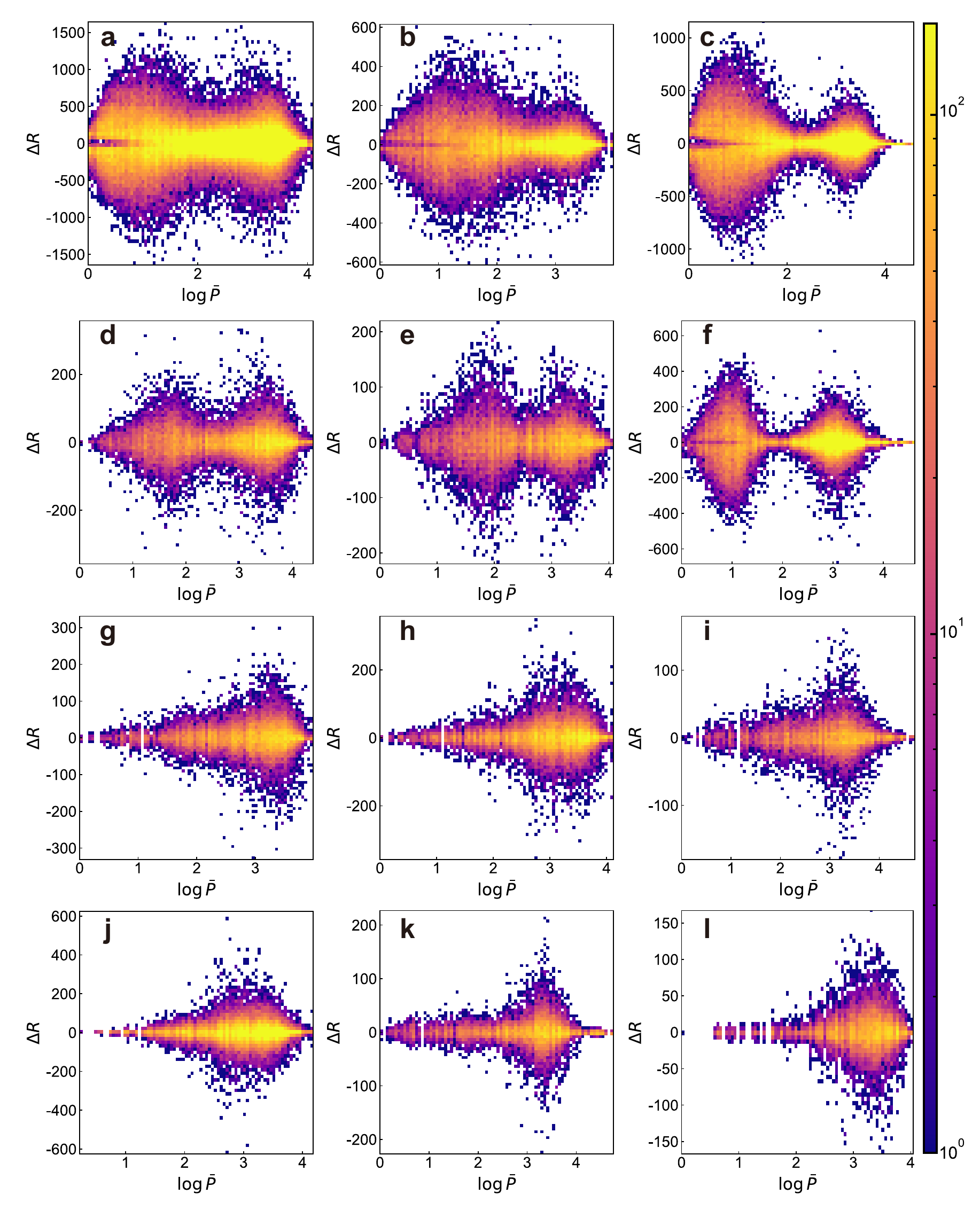}}
	\caption{{\bf Surface plots of $\Delta R$ in a function of $\bar{P}$ for twelve sampled cities.}
		(a) Beijing, (b) Tianjin, (c) Foshan, (d) Shijiazhuang, (e) Hohhot, (f) Quanzhou, (g) Taiyuan, (h) Nanchang, (i) Guiyang, (j) Changchun, (k) Nanning, and (l) Jinan.
	}
	\label{fig6}
\end{figure*}

\begin{figure*}
	\centering
	{\includegraphics[width=\linewidth]{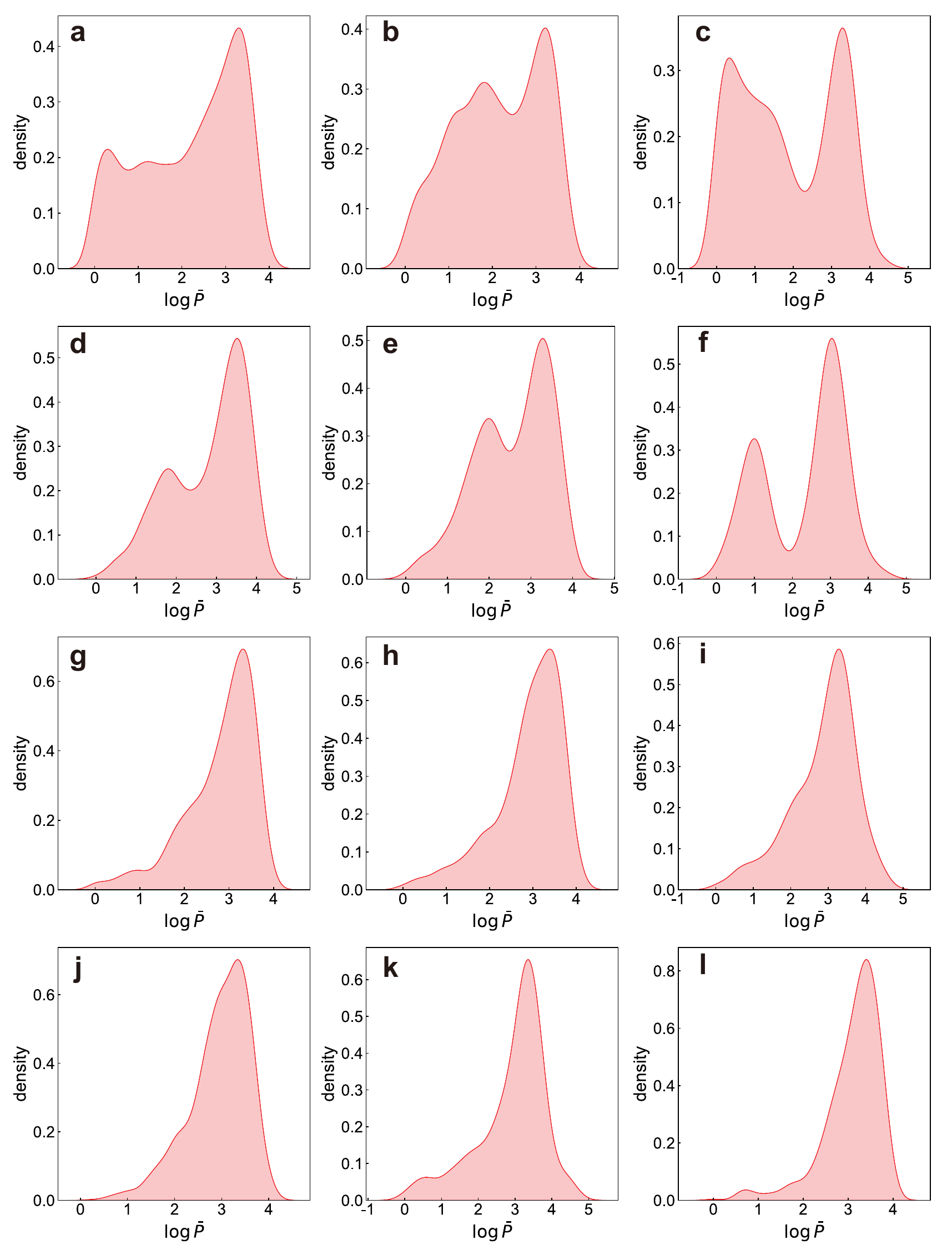}}
	\caption{{\bf The kernel density estimation plots of inflow distributions for twelve sampled cities.}
		(a) Beijing, (b) Tianjin, (c) Foshan, (d) Shijiazhuang, (e) Hohhot, (f) Quanzhou, (g) Taiyuan, (h) Nanchang, (i) Guiyang, (j) Changchun, (k) Nanning, and (l) Jinan.
	}
	\label{fig7}
\end{figure*}

\begin{figure*}
	\centering
	{\includegraphics[width=0.8\linewidth]{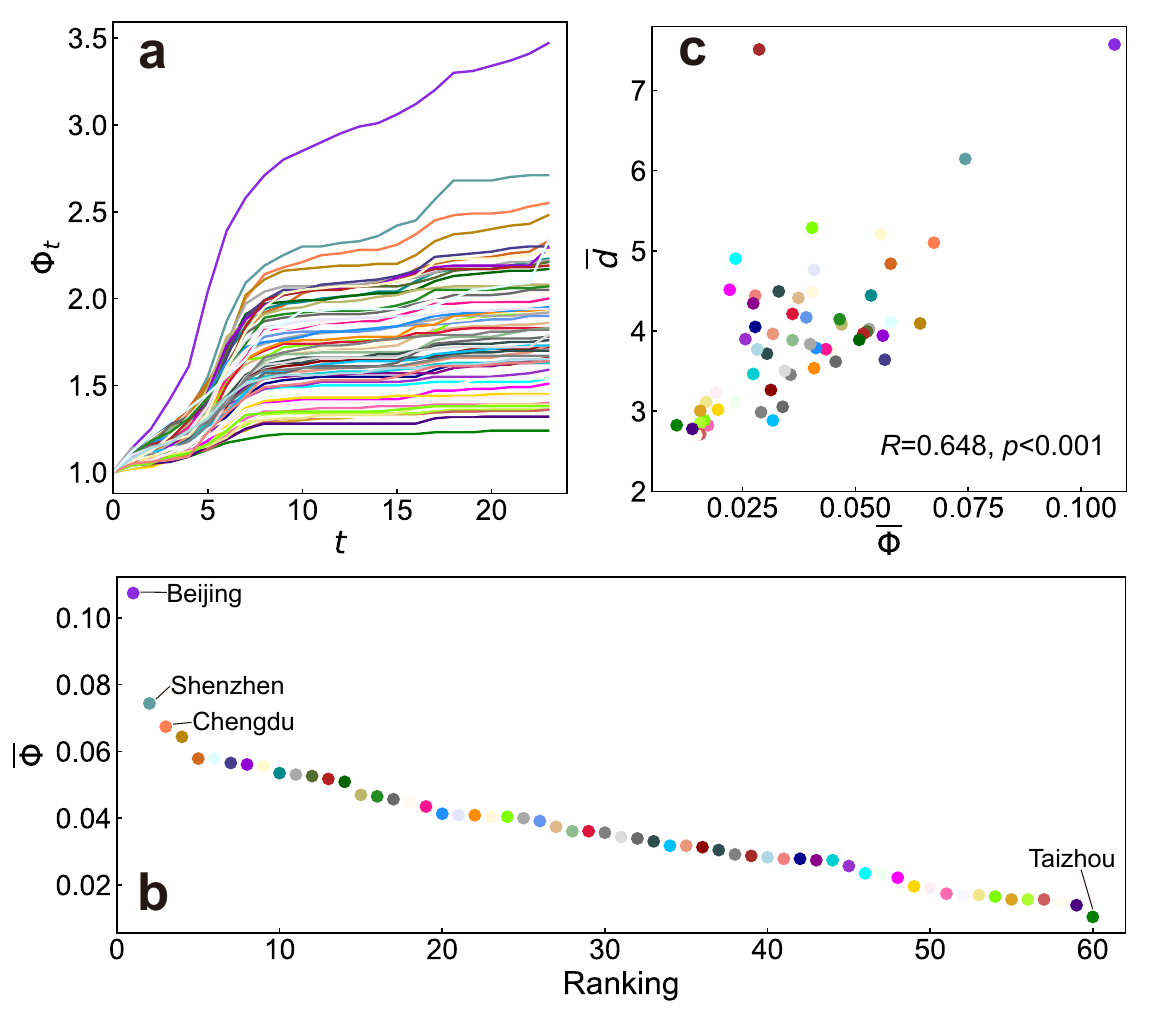}}
	\caption{{\bf Comparison of ranking dynamics across cities.}
		(a) Rank turnover $\Phi_t$ at hour $t$ for studied cities, defined as the number $N_t$ of locations ever seen in the ranking list up to $t$ relative to list size $N_0$. 
		(b) 60 cities ranked according to decreasing values of $\bar{\Phi}$. 
		(c) Correlation between the average turnover rate $\bar{\Phi}$ and the average daily travel distance $\bar{d}$. Each city is distinguished by a unique color.
	}
	\label{fig8}
\end{figure*}

\end{document}